\documentclass[aps,prd,preprintnumbers,nofootinbib,preprint]{revtex4}
\usepackage{bm}
\usepackage{latexsym}
\usepackage{dcolumn}
\usepackage{amsmath,amsfonts,amssymb}
\usepackage{graphicx,epsfig}
\usepackage{amsmath,amssymb}

\def\be {\begin{equation}}
\def\ee {\end{equation}}
\def\bea {\begin{eqnarray}}
\def\eea {\end{eqnarray}}
\def\bc {\begin{center}}
\def\ec {\end{center}}
\def\bfg {\begin{figure}}
\def\efg {\end{figure}}
\def\bi {\begin{itemize}}
\def\ei {\end{itemize}}

\begin{document}

\title{\textbf{Regular black hole metrics and the weak energy condition}}

\author{Leonardo Balart $^{1,2}$} \email[email: ]{leonardo.balart@ufrontera.cl}
\author{Elias C. Vagenas $^3$} \email[email: ]{elias.vagenas@ku.edu.kw}

\vspace{4ex}

\affiliation{$^1$~I.C.B. - Institut Carnot de Bourgogne UMR 5209
CNRS, Facult$\acute{e}$ des Sciences Mirande, Universit$\acute{e}$ de Bourgogne,  9 Avenue Alain Savary, BP 47870, 21078 Dijon Cedex, France}

\affiliation{$^2$~Departamento de Ciencias F$\acute{\imath}$sicas, Facultad de Ingenier$\acute{\imath}$a y Ciencias, 
Universidad de La Frontera, Casilla 54-D, Temuco, Chile}

\affiliation{$^3$~Theoretical Physics Group, Department of Physics, Kuwait University, 
P.O. Box 5969, Safat 13060, Kuwait}


\begin{abstract}
\par\noindent
In this work we construct a family of spherically symmetric, static, charged regular black hole metrics in the context of 
 Einstein-nonlinear electrodynamics theory. The construction of the charged regular black hole metrics is based 
on three requirements:   
 (a) the weak energy condition should be satisfied, (b) the energy-momentum tensor should have the symmetry 
$T^{0}_{0}=T^{1}_{1}$, and (c) these metrics have to asymptotically behave as the 
Reissner-Nordstr\"{o}m black hole metric. In addition, these charged regular black hole metrics  depend on two parameters which 
for specific values yield regular black hole metrics that already exist in the literature. 
Furthermore, by relaxing the third requirement, we construct more general 
regular black hole metrics which do not behave asymptotically as a Reissner-Nordstr\"{o}m black hole metric.
\end{abstract}
\maketitle
%
%
%
\section{Introduction}
\label{intro}
%
%
%
\par\noindent
Charged regular black hole solutions exist in the framework of 
 Einstein-nonlinear electrodynamics theory and are obtained as solutions of Einstein equations 
that are characterized by the fact that the metric as well as the curvature invariants 
$R$, $R_{\mu\nu} R^{\mu\nu}$, $R_{\kappa\lambda\mu\nu} R^{\kappa\lambda\mu\nu}$ 
do not present singularities anywhere \footnote{For a recent review see \cite{Ansoldi:2008jw}.}.
\par\noindent
The Bardeen black hole  is the first of a series of regular black hole solutions obtained \cite{bardeen}. 
If we write the most general form of a static line element with spherical symmetry
\begin{equation}
ds^2= -f(r) dt^2 + f(r)^{-1}dr^2 + r^2 (d \theta^2+\sin^2\theta d\phi^2)
\,\,\label{elem-intr} \, ,
\end{equation}
where the metric function can be written as
\begin{equation}
f(r) = 1 - \frac{2 m(r)}{r}
\,\,\label{f-intr} \, ,
\end{equation}
then, we can write the mass function of the Bardeen black hole as   
\cite{bardeen} \footnote{Henceforth, we use geometrized units, i.e., $G=c=1$.}
\begin{equation}
m(r) = \frac{M r^3}{(r^2 + g^2)^{3/2}}
\,\,\label{m-bard} \, .
\end{equation}
\par\noindent
Such a metric has event horizons located at $r_{\pm}$ if $g^2 \leq (16/27)M^2$, where $g$ can 
be interpreted as the monopole charge of a self-gravitating
magnetic field described by nonlinear electrodynamics~\cite{AyonBeato:2000zs}.
Furthermore, if $r \rightarrow \infty$, we get that the metric behaves as
\begin{equation}
f(r) \rightarrow 1 - \frac{2 M }{r} + \frac{3 M g^2}{r^3}   \,\,\label{lim-inf-Bard} \, .
\end{equation}
\par\noindent
Similarly, when $r \rightarrow 0$ the Bardeen metric function behaves as the de Sitter black hole, that is as
\begin{equation}
f(r) \rightarrow 1 - \frac{2 M }{g^3} r^2 \,\,\label{lim-0-Bard} \, .
\end{equation}
\par\noindent
Later on, other solutions~\cite{AyonBeato:1998ub}-\cite{Dymnikova:2004zc} were obtained, using the F-P dual formalism, 
by considering the action of general relativity coupled to nonlinear electrodynamics, namely
\begin{equation}
S =  \int d^4x \sqrt{-g} \left(\frac{1}{16 \pi}R - \frac{1}{4 \pi} L(F)\right)
\,\,\label{action-NL} \, . 
\end{equation}
\par\noindent
Here,  the Lagrangian $\emph{L}(F)$ is a nonlinear function of the Lorentz invariant 
$F = \frac{1}{4} F^{\mu\nu}F_{\mu\nu}$ which, for weak fields, describes the Maxwell theory, and the corresponding 
regular black hole solution asymptotically behaves as the Reissner-Nordstr\"{o}m black hole.
\par\noindent
The regular black hole may also be characterized by energy conditions \cite{Hawking:1973uf, Carroll:2004st} 
that the corresponding energy-momentum tensor  should satisfy.  In the context of regular black holes, 
three energy conditions have been utilized by several authors. In particular:
\begin{itemize}
\item
The strong energy condition (SEC) states that $T_{\mu\nu} t^\mu t^\nu \geq \frac{1}{2} 
T^\mu_{\, \, \, \mu} t^\nu t_\nu$ for all timelike vector $t^\mu$. This condition implies that gravitational force is attractive.
\item
The dominant energy condition (DEC) states that $T_{\mu\nu} t^\mu t^\nu \geq 0$ and that $T^{\mu\nu} t_\mu$ must 
be non-spacelike for all timelike vector $t^\mu$ or, equivalently, $T^{00}\geq |T^{i j}|$ for each $i, j =1, 2, 3$. 
This means that the local energy density measured by a given observer must be nonnegative, and that the speed 
of the energy density flow associated with this observer cannot exceed the speed of light.
\item
The weak energy condition (WEC) states that the energy density of matter measured by an
observer, whose 4-velocity is $t^\mu$, satisfies $T_{\mu\nu} t^\mu t^\nu \geq 0$ for all timelike vector $t^\mu$, that is,
the local energy density cannot be negative for all observers. The DEC implies the WEC.

\end{itemize}
\par\noindent
When a black hole is regular, the SEC is necessarily violated somewhere inside the horizon \cite{Zaslavskii:2010qz}.
However, a regular black hole could satisfy the WEC or the DEC everywhere \cite{Dymnikova:2004zc}.
\par\noindent
Considering the line element~(\ref{elem-intr}) and the metric function~(\ref{f-intr}), we can write the components of the respective
energy-momentum tensor as
\begin{equation}
T^0_{\, \, \, 0} = T^1_{\, \, \, 1} = \frac{2}{8 \pi r^2} \frac{d m(r)}{d r} \,\, , \,\,\,  T^2_{\, \, \, 2} = T^3_{\, \, \, 3}
=  \frac{1}{8 \pi r} \frac{d^2m(r)}{dr^2}  \,\,\label{tens-gral} \, ,
\end{equation}
and hence the WEC can be expressed equivalently in terms of the mass function by the following inequalities
\begin{equation}
\frac{1}{r^2} \frac{d m(r)}{d r} \geq 0
\,\,\label{1-wec-r} \, ,
\end{equation}
and
\begin{equation}
\frac{2}{r}\frac{dm(r)}{d r} \geq \frac{d^2m(r)}{dr^2} \,\,
\label{2-wec-r} \, .
\end{equation}
\par\noindent
From these relations, it is obvious that the Bardeen black hole satisfies the WEC everywhere \cite{bardeen}.
The same applies to regular black hole solutions reported in Refs. \cite{AyonBeato:1998ub},
\cite{AyonBeato:2004ih}, \cite{Dymnikova:2004zc}, \cite{Ansoldi:2006vg}, \cite{Dymnikova:1992ux}, 
\cite{Nicolini:2005vd}, and \cite{Hayward:2005gi}.
However, there are other regular black hole solutions reported in Refs.~\cite{AyonBeato:1999rg}, \cite{AyonBeato:1999ec},
\cite{AyonBeato:2004ih}, \cite{Bronnikov:2000vy}, and ~\cite{Burinskii:2002pz}, which do not satisfy the WEC.
\par\noindent
The regular black hole solutions that satisfy the WEC and their energy-momentum tensor has the symmetry 
$T^0_{\, \, \, 0} = T^1_{\, \, \, 1}$, necessarily have de Sitter behavior at $r \rightarrow 0$ as was shown in 
Ref. \cite{Dymnikova:1992ux}, and illustrated here in Eq.~(\ref{lim-0-Bard}) for 
the Bardeen black hole solution. As we will see later, this leaves an extra condition which allows the 
WEC to be better exploited compared to the DEC  when one is building solutions. 
However, it should be noted that a de Sitter behavior at the center of a regular black hole is not sufficient by itself  
to ensure that the solution satisfies the WEC.  In addition, it has been shown that  
 for a regular black hole in nonlinear electrodynamics which satisfies the WEC,  the non-existence theorems 
 \cite{Bronnikov:2000vy} can be circumvented by removing the condition of the Maxwell weak field limit imposed 
at the center of the black hole. In this way, regular black hole solutions with electric charge do exist \cite{Dymnikova:2004zc}. 

\par\noindent
Furthermore, there are other features that characterize regular black holes which are due to the nonlinearities of the 
field equations.  For example, the thermodynamic quantities of the regular black holes do not satisfy the Smarr 
formula \cite{Breton:2004qa}, the identity of Bose-Dadhich~\cite{Bose:1998uu} which refers to the relation between 
the Brown-York energy and the Komar charge, is not satisfied by regular black holes~\cite{Balart:2009xr}.

\par\noindent
In all the above-mentioned solutions, one asymptotically recovers the Schwarzschild black hole metric, and if the condition
$r^2 m'(r)\neq 0$ as $r\rightarrow \infty$ is satisfied, then one recovers the Reissner-Nordstr\"{o}m black hole metric.
 
\par\noindent
In this Letter,  in the context of  Einstein-nonlinear electrodynamics theory, we will construct a family of spherically symmetric, static, charged regular black hole metrics 
without utilizing the aforesaid methods but by imposing three conditions: (a) the weak energy condition should be satisfied, 
(b) the energy-momentum tensor should have the symmetry $T^{0}_{0}=T^{1}_{1}$, and 
(c) these metrics have to asymptotically behave as the 
Reissner-Nordstr\"{o}m black hole metric.  In addition, by relaxing the third condition, i.e., condition (c),  
we construct more general regular black hole metrics which do not behave asymptotically as a Reissner-Nordstr\"{o}m black hole metric.
The Letter is organized as follows. In Sec. II,  we present the equations that we will use in the construction of our metrics.
In Sec. III, we  obtain a general metric for charged regular black holes that satisfies the WEC and  asymptotically behaves as the
Reissner-Nordstr\"{o}m black hole metric. In addition, we  discuss a specific case of the aforesaid general metric function.
In Sec. IV, we extend the analysis to obtain metric functions that do not necessarily asymptotically behave as the
Reissner-Nordstr\"{o}m solution. Finally, in Sec. V, we briefly summarize our results.
%
%
%
\section{WEC equations}
\label{sec:2}
%
%
%
\par\noindent
Up to now, regular black holes solutions have been constructed by searching for, or postulating, the Lagrangian function $L(F)$  
in the framework  of F-P dual formalism  \cite{Salazar:1987ap}
\footnote{The F-P dual formalism is briefly presented in Appendix A.}.
As already mentioned in the Introduction, here 
we will construct a family of  spherically symmetric, static, charged regular black hole metrics 
by imposing three conditions: (a) the weak energy condition should be satisfied, 
(b) the energy-momentum tensor should have the symmetry $T^{0}_{0}=T^{1}_{1}$, and 
(c) these metrics have to asymptotically behave as the Reissner-Nordstr\"{o}m black hole metric. 
Therefore, in this section we will derive the equations which will be used in this construction, from the WEC.
\par\noindent
For simplicity and future convenience, we replace the variable $r$   that appears in Eqs. (\ref{1-wec-r}) 
and (\ref{2-wec-r}) with a new variable $x$  which is defined as $r=1/x$ and obviously $x \in [0,\infty)$.
It is evident that by employing the new variable $x$,  the WEC inequalities, namely Eqs. (\ref{1-wec-r}) 
and (\ref{2-wec-r}), now read
\begin{equation}
x^{4} \frac{d m(x)}{d x} \leq 0
\,\,\label{1-wec-x} \,
\end{equation}
and
\begin{equation}
4\, x^{3}  \frac{d m(x)}{d x} + x^{4} \frac{d^2m(x)}{dx^2} \leq 0 \,\,\label{2-wec-x} \, .
\end{equation}
\par\noindent
It should  be noted that these conditions plus regularity imply that the regular black hole metric function must 
satisfy the following limit
\begin{equation}
-\frac{d m(x)}{d x}x^4 \rightarrow b   \hspace{1cm}  \mbox{ when } \hspace{1cm} x \rightarrow \infty  \,\,\label{cond-imp} \, 
\end{equation}
where $b $ is a positive constant.
\par\noindent
At this point, we should stress that if we want the regular black hole metric to behave asymptotically 
as the Schwarzschild black hole, then we have to demand our metric function to satisfy the following condition

\begin{equation}
m(x) \neq 0  \hspace{1cm}  \mbox{ when } \hspace{1cm} x \rightarrow 0  \,\,\label{cond-schw} \, .
\end{equation}
\par\noindent
However, if we want the regular black hole metric to behave asymptotically as the Reissner-Nordstr\"{o}m black hole, 
then we must also require our metric function to satisfy the condition
\begin{equation}
-\frac{d m(x)}{d x} \neq 0   \hspace{1cm}  \mbox{ when } \hspace{1cm} x \rightarrow 0  \,\,\label{cond-reiss} \, .
\end{equation}
\par\noindent
It is easily seen that two mass functions which satisfy the aforesaid conditions, i.e. Eqs.~(\ref{1-wec-x})-(\ref{cond-reiss}), 
are defined by the following WEC equations
\begin{equation}
-\frac{d m(x)}{d x} = \frac{c_1}{(1 + c_2 \, x^{\alpha})^{4/\alpha}}  \,\,\label{deriv-func-1} \,
\end{equation}
and
\begin{equation}
-\frac{d m(x)}{d x} = \frac{c_3}{(1 + c_4 \, x^{1/\beta})^{4 \beta}}  \,\,\label{deriv-func-2} \,
\end{equation}
\par\noindent
where $\alpha$ is a  positive integer,  $\beta$ is a positive constant,  $c_1, c_2, c_3$, and $c_4$ are arbitrary 
but also positive constants  related by $c_1 = (c_2)^{4/\alpha} b$ and $c_3 = (c_4)^{4/\beta} b$. 
However, the latter mass function has to be discarded since its expansion leaves fractional powers (with the exception  
of the case with $\beta = 1$).
%
%
%
\section{Regular black hole metrics}
\label{sec:3}
%
%
%
\par\noindent
In this section, we will construct a family of spherically symmetric, static, charged  regular black hole metrics by
imposing the following three conditions on them: (a) the weak energy condition should be satisfied, 
(b) the energy-momentum tensor should have the symmetry $T^{0}_{0}=T^{1}_{1}$, and 
(c) the metrics have to asymptotically behave as the Reissner-Nordstr\"{o}m black hole metric.
\par\noindent
First we transform the WEC equation  given by Eq.~(\ref{deriv-func-1}) into the integral form
\begin{equation}
m(x) = \int_x^\infty   \frac{c_1}{(1 + c_2 \, y^{\alpha})^{4/\alpha}} dy    \,\,\label{integral-x} 
\end{equation}
and then we compute the above integral. The mass function is, thus, given by the expression
\begin{equation}
m(x) = \frac{c_1}{3(c_2)^{4/\alpha}}\frac{1}{x^3} \,\,
_{2}F_1\left(\frac{3}{\alpha},\frac{4}{\alpha};\frac{3+\alpha}{\alpha};-\frac{1}{c_2 \, x^\alpha}\right)  \,\,
\label{sol-x} 
\end{equation}
where $_{2}F_1(a,b;c;z)$ is the Gauss hypergeometric function .
\par\noindent
At this point, we demand the metric function given in Eq. (\ref{f-intr}), i.e.,
\begin{equation}
f(r) = 1 - \frac{2 m(r)}{r}~,
\label{f-intr-2}
\end{equation}
to behave asymptotically as the Reissner-Nordstr\"{o}m black hole metric, i.e.,
\begin{equation}
f(r) = 1 - \frac{2 M}{r} + \frac{q^2}{r^2} \,\,
\label{asymp} \, .
\end{equation}
Now, we substitute Eq. (\ref{sol-x}) in Eq. (\ref{f-intr-2}) and Taylor expand it around $r=0$.  
By comparing the coefficients of the series expansion at the asymptotic limit, i.e., $r\rightarrow\infty$, 
with the corresponding ones in Eq. (\ref{asymp}), we define the constants 
$c_{1}$ and $c_{2}$ as follows
\bea
c_1 &=& \frac{q^{2}}{2}\\
c_{2} &=& \left[\frac{q^{2} \Gamma(\frac{1}{\alpha})\Gamma(\frac{\alpha + 3}{\alpha})}
{6 M \Gamma(\frac{4}{\alpha})}\right]^{\alpha}~.
\label{c1c2}
\eea
\par\noindent
Therefore, the mass function given by Eq. (\ref{sol-x}) becomes
\begin{equation}
m(r) = \frac{r^3 q^2 }{6} \left(\frac{6 \, \Gamma(\frac{4}{\alpha})}{\Gamma(\frac{1}{\alpha})\Gamma(\frac{\alpha+3}{\alpha})}\frac{M}{q^2} \right)^4\,
_{2}F_1\left(\frac{3}{\alpha},\frac{4}{\alpha};\frac{3+\alpha}{\alpha};-\left(\frac{6 \, \Gamma(\frac{4}{\alpha})}{\Gamma(\frac{1}{\alpha})
\Gamma(\frac{\alpha+3}{\alpha})}\frac{M}{q^2} r \right)^\alpha\right)~.
\,\,\label{sol-r} 
\end{equation}
This is the mass function of a charged regular black hole metric given by Eq. (\ref{f-intr-2}) 
which asymptotically behaves as the Reissner-Nordstr\"{o}m black hole if  $\alpha$ is a positive constant. 
It is noteworthy that for $\alpha = 2$, we retrieve the regular black hole metric given in Ref.~\cite{Dymnikova:2004zc}.
\par\noindent
Furthermore, in the context of   Einstein-nonlinear electrodynamics theory,  
the electric field associated with the above regular black hole metric is given as \cite{leo+elias}
\begin{equation}
E = \frac{q}{r^2} \left(1 + \left(\frac{\Gamma(\frac{1}{\alpha})\Gamma(\frac{3+\alpha}{\alpha})}{6 \,
\Gamma(\frac{4}{\alpha})}\frac{q^2}{M r}\right)^\alpha \right)^{-(4+\alpha)/\alpha}
\,\,\label{E-F} \, ,
\end{equation}
which behaves as $E = \frac{q}{r^2}$ when $r \rightarrow \infty$.
\par\noindent
Finally,  as an example, we choose $\alpha = 3$ in which case the metric function is of the form
\begin{equation}
f(r) = 1 - \frac{2 M}{r} \left(1-\frac{1}{\left(1+\left(\frac{2 M r}{q^2}\right)^3\right)^{1/3}}\right)\,\,\label{alp-3} \, .
\end{equation}
\par\noindent
This regular black hole metric has event horizons if the electric charge satisfies the condition $q \leq 1.0257 M$.
Moreover, in the context of   Einstein-nonlinear electrodynamics theory, the associated electric field is given as 
 \cite{leo+elias}
\begin{equation}
E = \frac{q}{r^2} \left(1 + \left(\frac{q^2}{2 M r}\right)^3\right)^{-7/3} \,\,\label{E-F-alp-3} \, .
\end{equation}
%
%
%
\section{More General Metrics}
\label{sec:4}
%
%
%
\par\noindent
In this section, we will construct more general regular black hole metrics. 
The same analysis as in the previous section will be adopted here but we will not demand the metric to 
behave asymptotically as the Reissner-Nordstr\"{o}m black hole metric. For this reason, we will relax the condition given 
in Eq. (\ref{cond-reiss}).
 The mass function $m(x)$ will now satisfy the following WEC equation
\begin{equation}
-\frac{d m(x)}{d x} = \frac{c_{1}x^{\mu - 4}}{(1 + c_{2}x^{\alpha})^{\mu/\alpha}}  \,\,\label{deriv-func-gral} 
\end{equation}
\par\noindent
where $\alpha$ and $\mu$ are integers and also $\alpha \geq 1$ and $\mu \geq 4$ and, thus, 
it is given by the expression
\begin{equation}
m(x) = \frac{c_{1}}{3 (c_{2})^{\mu / \alpha}}\frac{1}{ x^3} \,
 _{2}F_1\left(\frac{3}{\alpha},\frac{\mu}{\alpha};\frac{3+\alpha}{\alpha};
-\frac{1}{c_2 \, x^\alpha}\right)  \,\,\label{sol-x-gral} \, .
\end{equation}
The coefficients $c_{1}$ and $c_{2}$ are now given by
\bea
c_1 &=& \frac{q^{2}}{2}\frac{\Gamma(\frac{1}{\alpha})\Gamma(\frac{\mu}{\alpha})}
{{\Gamma(\frac{4}{\alpha})\Gamma(\frac{\mu - 3}{\alpha})}}\left(  \frac{6 \Gamma{(\frac{4}{\alpha})}}
{\Gamma(\frac{1}{\alpha})\Gamma(\frac{\alpha+3}{\alpha})} \frac{M}{q^{2}}\right)^{4-\mu}\\
c_{2} &=& \left[\frac{6 \Gamma(\frac{4}{\alpha})}
{\Gamma(\frac{1}{\alpha})\Gamma(\frac{\alpha + 3}{\alpha})})\frac{M}{q^{2}}\right]^{-\alpha}~.
\label{c1c2_gen}
\eea
\par\noindent
Thus,  the mass function of the regular black hole metric becomes
\begin{eqnarray}
m(r) &=& \frac{r^3 q^2 }{6} \frac{\Gamma(\frac{1}{\alpha})\Gamma(\frac{\mu}{\alpha})}
{\Gamma(\frac{4}{\alpha})\Gamma(\frac{\mu-3}{\alpha})}
\left(\frac{6 \, \Gamma(\frac{4}{\alpha})}{\Gamma(\frac{1}{\alpha})\Gamma(\frac{\alpha+3}{\alpha})}
\frac{M}{q^2} \right)^{4}\,
\nonumber \\
&& _{2}F_1\left(\frac{3}{\alpha},\frac{\mu}{\alpha};\frac{3+\alpha}{\alpha};-\left(\frac{6 \, \Gamma(\frac{4}{\alpha})}
{\Gamma(\frac{1}{\alpha})\Gamma(\frac{\alpha+3}{\alpha})}\frac{M}{q^2} r \right)^\alpha\right)  \,\,
\label{sol-r-gral} ~.
\end{eqnarray}
This is the mass function of a charged regular black metric given by Eq. (\ref{f-intr-2}),  but  
which does not asymptotically behave as the Reissner-Nordstr\"{o}m black hole metric  except for the case with  $\mu = 4$.
\par\noindent
It is worthy of notice that, with an appropriate choice of parameters, we can derive from Eq.~(\ref{sol-r-gral}) 
several regular black hole metrics which already exist in the literature. For instance, if we choose 
$\mu = 5$ and $\alpha = 2$, we obtain
\begin{equation}
m(r) = \frac{M r^3}{(r^2 + \frac{\pi^2 q^4}{64 M^2})^{3/2}} \,\,\label{bard-h} ~.
\end{equation}
By replacing $\pi^2 q^4/(64 M^2)$  with  $g^2 $ in Eq.~(\ref{bard-h}) , we recover the Bardeen metric
whose mass function is given by Eq.~(\ref{m-bard}).
\par\noindent
Now, if we take $\mu = 6$ and $\alpha = 3$, and we replace the factor $\frac{q^6}{8 M^3}$ with $2 l^2 M$, 
we obtain the regular black hole metric given in Ref.~\cite{Hayward:2005gi} with mass function
\begin{equation}
m(r) = \frac{M r^3}{r^3 + 2 l^2 M} \,\,\label{hayw} \, .
\end{equation}
\par\noindent
Finally, if we choose $\mu = 3$ and let $\alpha$  be arbitrary, then we obtain the following mass function
\begin{equation}
m(r) = M \left(1 - \frac{1}{\left(1 + \left(\frac{2 M}{q^2} r\right)^3\right)^{(\alpha - 3)/3}}\right) \,\,\label{dag} \, .
\end{equation}
%
%
%
\section{Conclusions}
\label{sec:10}
%
%
%
%
\par\noindent
In this Letter, we have constructed a family of spherically symmetric, static, charged regular black hole metrics 
in the context of  Einstein-nonlinear electrodynamics theory. Our analysis is based on the fact that we 
impose three conditions on the black hole metrics:  
(a) the weak energy condition should be satisfied, (b) the energy-momentum tensor should have the symmetry 
$T^{0}_{0}=T^{1}_{1}$, and (c) these metrics have to asymptotically behave as the 
Reissner-Nordstr\"{o}m black hole metric.
Moreover, by relaxing the third requirement, we construct more general 
regular black hole metrics which do not behave asymptotically as a Reissner-Nordstr\"{o}m black hole metric. 
In addition, we  discuss as examples several special cases of the more 
general regular black hole metrics. These special cases have been obtained by choosing specific values for the parameters 
that characterize the mass function of the more general regular black hole metric.
Some of these regular black hole metrics already exist in the literature but they are obtained in the context 
of F-P dual formalism. All the above regular black hole  metrics also satisfy the DEC, 
although it was not imposed as a condition.
%
%
%
\par\noindent\\
{\bf Acknowledgments}\\
%
%
L.B. would like to thank Direcci\'on de Investigaci\'on y Postgrado de la Universidad 
de La Frontera (DIUFRO) for the financial support.
%
%
\par\noindent\\
{\bf Appendix A: Dual P Formalism}\\
%
%
An equivalent method for deriving the regular black hole metrics obtained here is the F-P dual formalism. 
For this reason, we briefly present here the description based on the F-P dual representation of nonlinear electrodynamics obtained by
a Legendre transformation~\cite{Salazar:1987ap} and reproduce the results of Sec. III.
\par\noindent
The regular black hole metrics can be described by the metric function and its corresponding electromagnetic field 
which arise as a solution of Einstein field equations coupled to a nonlinear electrodynamics, that is of the action given 
by Eq.~(\ref{action-NL}).
One can also describe the considered system in terms of an auxiliary field defined by
$P_{\mu\nu} = (dL/dF) F_{\mu\nu}$. The dual representation is obtained by means of a Legendre transformation
\begin{equation}
H =  2 F \frac{d L}{d F}- L \,\,\label{Legendre} \,
\end{equation}
which is a function of the invariant $P = \frac{1}{4}P_{\mu\nu}P^{\mu\nu}$. 
Thus, we can express the Lagrangian $L$ depending on $P_{\mu\nu}$ as
\begin{equation}
L =  2 P \frac{d H}{d P} - H \,\,\label{Lag} \, ,
\end{equation}
and the electromagnetic field as
\begin{equation}
F_{\mu\nu} = \frac{d H}{d P} P_{\mu\nu} \,\,\label{fmunu} \, .
\end{equation}
\par\noindent
The energy-momentum tensor in the F-P dual representation is given by
\begin{equation}
T_{\mu\nu} = \frac{1}{4 \pi}\frac{d H}{d P} P_{\mu\alpha}P_\nu^\alpha - \frac{1}{4 \pi} g_{\mu\nu} \left(2 P \frac{d H}{d P} - H \right)
\,\,\label{tensormunu} \, .
\end{equation}
\par\noindent
It follows from the components of $T_{\mu\nu}$ that $M'(r) = - r^2 H(P)$. 
Hence, we can obtain the corresponding mass function.
\par\noindent
As an example, we give the function $H(P)$ for the regular black hole metrics of Sec. III
\begin{equation}
H(P) = \frac{P}{(1 + \Omega \, (-P)^{\alpha/4})^{4/\alpha}}
\,\,\label{H-gral} \,
\end{equation}
where
\begin{equation}
P = -\frac{q^2}{2 r^4}
\,\,\label{P-inva} \,
\end{equation}
and $\Omega$ is defined as
\begin{equation}
\Omega = \left(\frac{q^{3/2}}{6 M}\frac{\Gamma(\frac{1}{\alpha}) \Gamma(\frac{3+\alpha}{\alpha})}{\Gamma(\frac{4}{\alpha})}\right)^\alpha
\,\,\label{Omega-def} \, .
 \end{equation}
%
%
%
%
%

\end{document}